\documentclass{PoS}
\usepackage{epsfig}

\PoS{PoS(LAT2005)186}

\title{Transport Coefficients of Gluon Plasma from Lattice QCD 
\footnote{Revised Version of Lattice 05 Proceedings,   YGHP-05-36}}

\ShortTitle{Transport Coefficient of Gluon Plasma from Lattice QCD}

\author{\speaker{Sunao Sakai}
    \\

        Yamagata University\\
        E-mail: \email{sakai@e.yamagata-u.ac.jp}}

\author{Atsushi Nakamura\\

        RIISE Hiroshima University\\
        E-mail: \email{nakamura@riise.hiroshima-u.ac.jp}}

\abstract{
\indent
RHIC is now confirming the discovery of the ``New State
of Matter'', whose properties are gradually revealed by experimental
data.
It is not the free gas state, but well described by a fluid with very
small viscosity.
It is now highly desired to calculate the value of the viscosity from
the fundamental theory, i.e.,  QCD just above the transition
temperature. 
In this report we present our
calculation of the transport coefficient of gluon system
on $24^3\times 8$ lattice in the
quench approximation.  Simulations are carried out in the range,
$1.4 \le T/T_c \le 24$. 
In the temperature region slightly
above the transition,  where the perturbative calculation
is not applicable,  the shear viscosity($\eta$) is smaller than
typical hadron masses.  The bulk viscosity is consistent
with zero within the range of error bars in  $1.4 \le T/T_c \le 24$.
We compare our results with the
perturbative calculations in large $T/T_c$ region. It
is found that the lattice and perturbative results are 
consistent with each other there.  
The ratio $\eta/s$ is around  $0.1-0.4$ in $T/T_c < 3$
region and satisfies the KSS bound\cite{KSS}.
In order to estimate the contribution from
high frequency part of the spectral function, we study the effects of 
a term $\rho^{high}$ 
proposed by Aarts and Resco\cite{Aarts}. It is found that until the
threshold mass becomes 
small, its effect is quite small, and that
viscosity decreases as the threshold decreases.
From these studies we 
think that 
although our result is obtained under an assumptions for the spectral
function, it gives a reasonable estimation for $\eta$($=\pi
d\rho/d\omega$ at
$\omega=0$), and qualitative results will not be changed when 
the accurate spectral function is obtained.

}

\FullConference{XXIIIrd International Symposium on Lattice Field
Theory\\

 25-30 July 2005\\

  Trinity College, Dublin, Ireland}

\begin{document}

\section{Introduction}
 The RHIC data are now revealing some properties of a ``New State of
Matter'', which is expected to be quark-gluon plasma(QGP). The jet
quenching data
suggest that its mean free path is short, and 
 the phenomenological analysis of the elliptic flow shows that, it could
be explained by a fluid model with small viscosity. Namely it is
almost perfect fluid. 
As the viscosity is proportional to mean free path\cite{Sakagami}, this
analysis also indicates that mean free path is short.
The mean free path $\lambda$ is related to scattering cross section
$\sigma$ and the number density $n$ as $\lambda \propto 1/(\sigma n)$.
Therefore the  small $\lambda$, means that the
``New State of Matter'' is a strongly interacting system.

It is a natural anticipation that transport coefficients of the QGP
must be smaller than an ordinary matter, like water or oil etc., because
QCD coupling in the deconfined phase will be larger than the
electromagnetic coupling constant $1/137$,
 but its value must
be calculated from fundamental theory of QCD.
Therefore it is now urgent to 
calculate transport coefficients by fully taking into account
non-perturbative effects, because
QGP near the transition temperature will be strongly interacting
In Ref.\cite{nakamura}
we published our results from lattice simulations in the $1.5 <
T/T_c< 2.1$ region. 
In this report, the study is extended to higher temperature region,
$T/T_c < 24$, and we compare our
results with perturbative calculations there. We also discuss how our
results depend on the
spectral function of the green function in high
frequency region\cite{Aarts}.

On a lattice, calculation of the transport
coefficients is formulated in the framework of the linear response
theory\cite{Zubarev,Horsley}.
\begin{eqnarray}
\eta 
= - \int d^{3}x' \int_{-\infty}^{t} dt_{1} e^{\epsilon(t_{1}-t)}
     \int_{-\infty}^{t_{1}}
     dt'<T_{12}(\vec{x},t)T_{12}(\vec{x'},t')>_{ret}
\label{off-diagonal}
\end{eqnarray}
\begin{eqnarray}
\frac{4}{3}\eta+\zeta
 = -\int d^{3}x' \int_{-\infty}^{t} dt_{1} e^{\epsilon(t_{1}-t)}
 \int_{-\infty}^{t_{1}}dt'
  <T_{11}(\vec{x},t)T_{11}(\vec{x'},t')>_{ret}
\label{diagonal}
\end{eqnarray}
where $\eta$ is shear viscosity, and  $\zeta$,  bulk viscosity.
$<T_{\mu \nu} T_{\rho \sigma}>_{ret}$ is a retarded Green's function
of energy momentum tensor at a given temperature. 
In the quenched model, $T_{\mu \nu}$
is written by the field strength tensor as follows.
\begin{eqnarray}
T_{\mu\nu}= 2Tr[F_{\mu\sigma}F_{\nu\sigma}
-\frac{1}{4}\delta_{\mu\nu}F_{\rho\sigma}F_{\rho\sigma}]
\end{eqnarray}
The field strength tensor is defined by the plaquette operator,
$U_{{ } \mu\nu}(x) = \exp{(ia^2 g F_{\mu \nu}(x))}$.

  The shear viscosity in Eq.\ref{off-diagonal} is also expressed 
by using a spectral function $\rho$ of the retarded Green's 
function $\rho(\omega)$\cite{Horsley} as follows.
\begin{eqnarray}
\eta = \pi \displaystyle{\lim_{\omega \rightarrow 0}}
 \frac{\rho(\omega)}{\omega}
= \pi\lim_{\omega \rightarrow 0} \frac{d \rho(\omega)}{d \omega}
\label{diff}
\end{eqnarray}

For the determination of the spectral function $\rho(\omega)$,
we use a well known fact that the spectral function of the retarded
Green's function at temperature $T$ is same as that of Matsubara-Green's
function. 
 Therefore our target is to calculate  Matsubara-Green's
function($G_{\beta}(t_n)$) on a lattice and determine $\rho(\omega)$ from it.

\section{Lattice calculation of $\eta$ from Matsubara-Green's function
 $G_{\beta}(t)$ }

Using the spectral function $\rho(\omega)$, 
$G_{\beta}(t)$ is expressed as follows 
\begin{eqnarray}
G_{\beta}(t) = 
-\frac{1}{\beta}\sum_{n}e^{-iw_n t} \int_{-\infty}^{\infty}
 \frac{\rho(\omega)}{i\omega_n-\omega} d\omega
= \int_{0}^{\infty} 
\frac{cosh(\omega(t-\beta/2))}{sinh(\omega \beta/2)}
 \rho(\omega) d\omega,
\label{f-series}
\end{eqnarray}
where $\omega_{n} = 2 \pi n/\beta$. Note that the last equality holds 
for the region $ 0 < t  <\beta(N_T)$: at $t=0$ and $\beta(=N_T)$ it could
not be used.

 In order to determine the spectral function $\rho(\omega)$ from
$G_{\beta}(t_n)$, which is given at discrete points $t_n=1,2...,N_T-1$
on a lattice, the most promising method
may be the maximum entropy method(MEM). However to determine the
$\rho(\omega)$ precisely by MEM, $G_{\beta}(t_n)$ on  $N_T > 30$ lattice
with high  statistics is necessary. 
However, we realized that $G_{\beta}(t_n)$ is very noisy\cite{sakai},
and we decided to start with a
relatively smaller lattice of large statistics
with some assumption for the spectral
function $\rho(\omega)$. 
 
The simplest non trivial form is a Bright-Wigner type proposed by
 Karsch and  Wyld\cite{Karsch}.
\begin{eqnarray}
\rho_{BW}(\omega)
 = \frac{A}{\pi}(\frac{\gamma}{(m-\omega)^2+\gamma^2}-
 \frac{\gamma}{(m+\omega)^2+\gamma^2}) .
\label{rho_ka}
\end{eqnarray}
In this case, shear viscosity is given by $ \eta = 4A\gamma
m/(\gamma^2+m^2)^2$, similarly for bulk viscosity $\zeta$.
As this formula already has 3 parameters, in order to determine them by
least squares, the lattice size in temperature direction($N_T$) must be
$N_T \geq 8$. Then the minimum lattice size is $24^3 \times 8$,
to obtain a non trivial results.

\begin{figure}
\epsfig{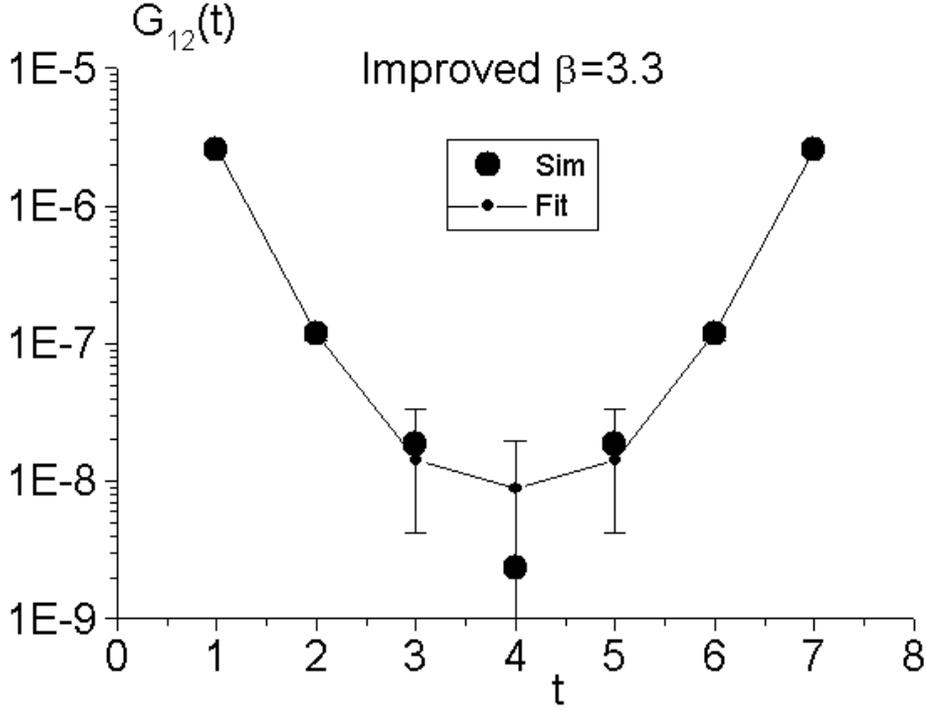}
\caption{Matsubara-Green's function $G_{12}(t)$ at $\beta=3.3$
}
\label{g12sim}
\end{figure}

Simulations are carried out by using the Iwasaki's improved action and
standard Wilson action.
The simulation points are at $\beta=$3.05,
3.3, 4.5 and 5.5 for the improved action and $\beta$=7.5 and 8.5 for Wilson
action.  With roughly $10^6$ MC measurements at each $\beta$, we
determine the Matsubara-Green's functions $G_{\beta}(t_n)$ as shown in
Fig. \ref{g12sim}.

Although the errors of the Matsubara-Green's function are still not
small in large $t$
region,
we can fit them with the spectral function $\rho$, given by
Eq.\ref{rho_ka}, using a least square package SALS. The fitting range is
$ 1 \le t \le 4$.
The viscosities are calculated from these parameters, 
and the errors are estimated by the jackknife method. The bin size of the
jackknife analysis is taken to be $10^5$ and $2\times 10^5$ for the improved
and Wilson actions, respectively.

\begin{figure}
\epsfig{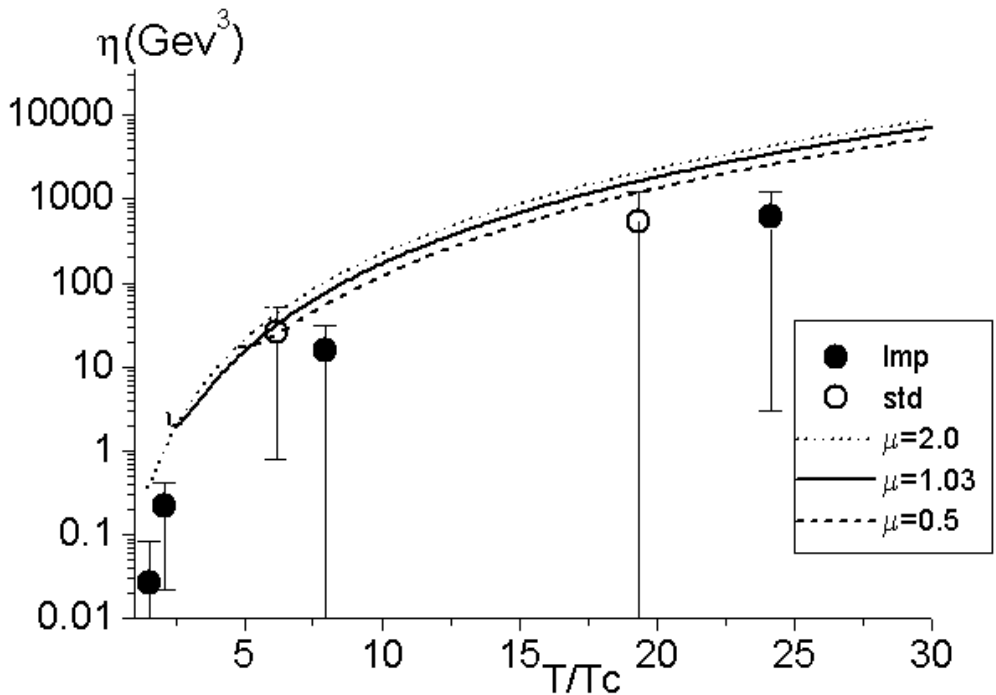}
\caption{Shear viscosity in physical unit from lattice(circles) and
perturbative calculations(lines). The lines show the ambiguity due to
 the
choice of  the scale parameter in two-loop running coupling formula.  }
\label{eta_phys}
\end{figure}

The bulk viscosity is equal to zero within the range of error bars,
while the shear viscosity remains finite.
In order to determine the shear viscosity in physical unit, the lattice
spacing $a(\beta)$ must be determined,
because, in the lattice calculation, $\eta \times a^3$ is obtained. 
The relation between lattice spacing $a$ and $\beta$ is studied
phenomenologically in $2.2 \le \beta \le
3.8$, and $ 5.58 \le \beta \le 6.5$ regions for improved\cite{Okamoto}
and
Wilson actions\cite{Edwards}, respectively. In the  larger $\beta$
region,
we adopt 2-loop renormalization group relation,
\begin{eqnarray}
 a(\beta)= \frac{1}{\Lambda} (\frac{6b_0}{\beta})^{-\frac{b_1}{2b_0^2}}
           \exp(\frac{-\beta}{12b_0}),
\label{a_inv}
\end{eqnarray}
with $b_0=33/(48\pi^2)$, and $b_1=(34/3)(3/16\pi^2)^2$. The $\Lambda$ is
determined by the $a(\beta)$ at
$\beta=$3.8 and 6.5, and they result in 
$T_c/\Lambda=$ 1.26 and 47.35 for the improved and Wilson
actions, respectively, where $T_c=270$MeV in the quenched model. 
 The results for the shear viscosity are shown in Fig. \ref{eta_phys}
with circle.
It is seen that $\eta$ from both actions are consistent with each
other. 
\section{Comparison with perturbative results}
As we have calculated $\eta$ at rather high temperature region,
we compare our results with the perturbative results. 
The perturbative results are summarized as
follows. The bulk viscosity $\zeta$ is zero \cite{Horsley,Kajantie}.
The shear viscosity in the next-to-leading log
is given by\cite{Arnold},
\begin{eqnarray}
 \eta = \frac{\eta_1 \cdot T^{3}}{g^{4}(\ln(\mu^{*}/g T)},
\end{eqnarray}
where $\eta_1=27.126$ and $\mu^{*}/T=2.765$, and $g$ is the
running coupling constant. We use 2-loop formula for it,
\begin{eqnarray}
 g^{-2}(T) = 2b_0 \ln(\mu \frac{T}{T_c}) +
           \frac{b_1}{b_0}\ln\left(2 \ln(\mu \frac{T}{T_c}) \right)
\label{g2}
\end{eqnarray}
\indent
Note that  the perturbative approximations are 
renormalization-scale $\mu$
dependent. The simplest choice would be $\mu=T_c/\Lambda_{QCD} \sim
1.03$\cite{Blaizot}.  However in order to see the ambiguity of the
perturbative
result due to $\mu$, we change $\mu$ in the range $0.5 \le \mu \le
2$. The results are also shown in Fig. \ref{eta_phys} with line. 
In the large $T$ region, the $\mu$ dependence is not large.
However at smallest $T$ of each line, it shows a small increase. It
means the break down of the perturbative calculation. However until the 
break down starts the agreement of the perturbative and lattice results
are satisfactory. 
From this, we think that although our result depends on the
assumption of the $\rho_{BW}$ given in Eq. \ref{rho_ka}, it may be a
reasonable approximation for $d\rho/d\omega$ at $\omega=0$.

\begin{figure}
\epsfig{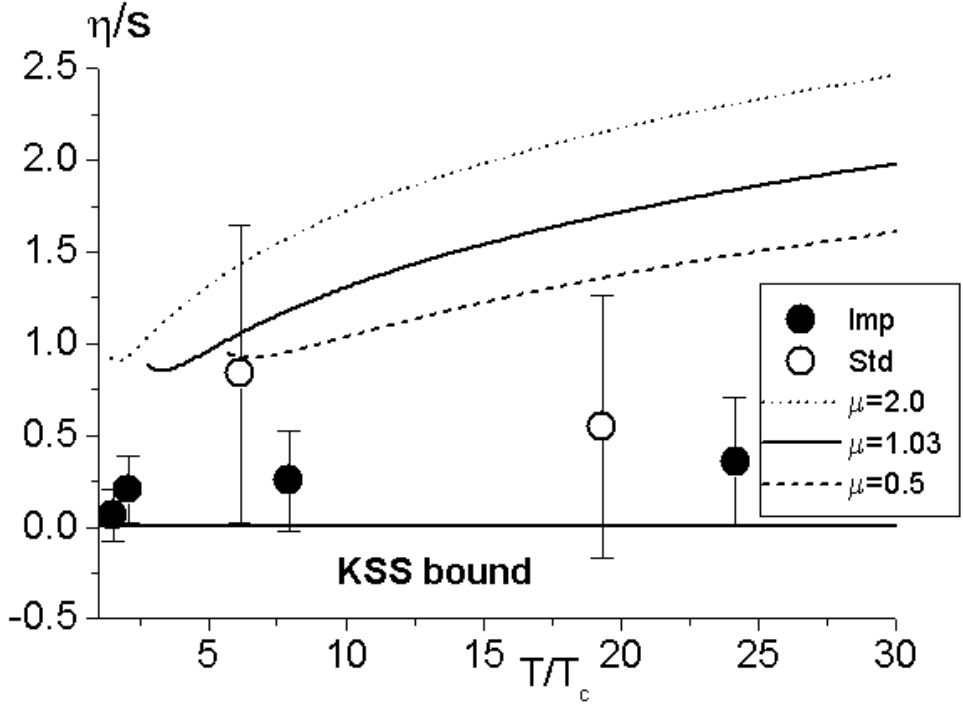}
\caption{$\eta/s$ obtained by lattice simulations(circles) and
 perturbative calculations(lines).
The $\mu$ is a scale parameter in the running coupling constant of Eq.3.2}
\label{ETAVSS}
\end{figure}

Let us proceed to the study of the $\eta/s$ ratio, recently discussed in
\cite{SON,KSS}.
The entropy on the lattice is reported by \cite{Okamoto}
and \cite{Boyd} in $T/T_c < 4.5$, for improved and Wilson actions
respectively.
In the higher temperature region, where the data are not available,
we use the high temperature limit values on $N_T=8$ lattice.
The results are shown in Fig. \ref{ETAVSS} with circles.
For the perturbative result on entropy, 
we use the formula of hard thermal loop calculation up to order
$g^3$\cite{Blaizot}. The results are also shown in Fig.\ref{ETAVSS} with
lines.\\
\indent
As explained in the Fig.\ref{eta_phys}, the slight increase at the small
$T$ region is observed for the perturbative results, indicating
breakdown of perturbative calculations.
The agreement between these two calculations are not good enough.
However we do not think that it is a real difficulty. 
Because if we decrease the scale parameter $\mu$, the agreement is
improved.
An important information from these results is the qualitative magnitude
of the ratio $\eta/s$.
In the $T/Tc < 3$ region, the ratio is 
small (0.1-0.4), where the perturbative results diverges. 
And for all the temperature regions  we have studied, it is less than
one. 
We don't think that it would become 10 times of present value, when the
accurate determination of the spectral function is carried out.  

\section{Discussions and conclusions}
\subsection{Fit of $G_{\beta}$ by other spectral functions }
 Aarts and Resco have proposed an another form for the
spectral function $\rho$\cite{Aarts},

\begin{eqnarray}
 \rho(\omega)=\rho^{low}(\omega) + \rho^{high}(\omega)
\label{rho_a}
\end{eqnarray}
\begin{eqnarray}
\frac{ \rho^{low}(\omega)}{T^4}=
 x \frac{b_1+b_2 x^2+...}{1+c_1 x^2+c_2 x^4+.. },
 \hspace{1.0cm} x=\frac{\omega}{T}
\label{rho_low}
\end{eqnarray}
\begin{eqnarray}
\rho^{high}(\omega)=\theta(\omega-2m_{th}) \frac{d_A (\omega^2-4m_{th}^2)^{5/2}}
{80 \pi^2 \omega }[n(\omega/2)+0.5]
\label{rho_high}
\end{eqnarray} 
where $d_A=N_c^2-1$, and $n(\omega)=1/(\exp(\omega/T)-1)$.
We consider three parameters case,
$ b_1$, $c_1$ and $m_{th}$.  The fit by SALS could be
made but $c_{1} < 0$.

We would like to remember that
the spectral function $\rho$ must satisfy the sum rule\cite{LeBellac},
\begin{eqnarray}
 \int_{-\infty}^{\infty} \frac{1}{2 \pi} \omega \rho(\omega) d\omega =1.
\end{eqnarray}
In the full parameter representation,  the cancellation of the 
divergence between 
$\rho^{low}$ and $\rho^{high}$ in high $\omega$ region results in
the convergence of the integration. But
in this three parameter form, the $\rho$ 
could not satisfy the sum rule. Then some modification is necessary.
But in this work we applied the unmodified form. Because 
the integration over $\omega$ in Eq.\ref{f-series} converges due to
the term $cosh(\omega(t-\beta/2))/sinh(\omega \beta/2)$, except for
$t=0$ and $t=\beta$, we think that phenomenologically, a possible 
contribution from high frequency part of $\rho(\omega)$ could be 
studied in this form.

As the another trial, we study effects of $\rho^{high}$ for the 
shear viscosity $\eta$. We assume $\rho$ is given by 
$\rho=\rho_{BW}+\rho^{high}$, where $\rho_{BW}$ is given by
Eq. \ref{rho_ka}.
By changing $m_{th}$, the change of  $\eta$ is studied at $\beta=3.3$ for the
improved action. When $\rho^{high}$ is absent ($m_{th}=\infty$), $\eta
a^3$=0.00225(201). If $m_{th}$ is set to be 5.0, 3.0 and 2.0, $\eta a^3$
becomes
0.00223(0.00191), 0.00194(0.00194) and 0.00126(0.00204), respectively.
And at $m_{th}=1.8$, the contribution from $\rho^{high}$ becomes larger than
the $G_{\beta}(t_n)$ of simulation at $t=1$, that fit could  not be done.
Generally as $m_{th}$ decreases the contribution from $\rho^{high}$ increases
and the $\rho$ in the small $\omega$ region is suppressed. In this case, 
it results in the decrease of $\eta$. 

  In this case, the $\rho_{BW}$ (without $\rho^{high}$), determined by 
SALS to fit the Matsubara  Green's function, satisfies the relation, 
\begin{eqnarray}
 \int_{-\infty}^{\infty} \frac{1}{2 \pi} \omega \rho_{BW}(\omega)
 d\omega < 1. 
\label{sum_ka}
\end{eqnarray}
As a full spectral function $\rho$ will have contributions from
scattering states etc., in addition to Bright-Wigner
type, Eq.\ref{sum_ka} does not mean that the full spectral
function violates the sum rule. 
 
 


\subsection{Conclusion and further studies}
On $24^3 \times 8$ lattice, we have calculated the Matsubara-Green's
function
and determine the transport coefficients of gluon plasma.
The bulk viscosity is consistent with zero, 
while shear viscosity remains finite, within the error bars of present
statistics.
In the high
temperature region, the agreement of the lattice and perturbative
calculation is satisfactory for the shear viscosity. The lattice result
on ratio $\eta/s$ in
$T/T_c \le 3$ is smaller than the extrapolation of the perturbative
calculation and satisfies the KSS bound.

Although our results depend on the
form of the spectral function
$\rho_{BW}$ given by Eq.\ref{rho_ka}, we think that the qualitative
features will
not be changed.  Because as discussed in the previous subsection, our
results are stable
if the high frequency part of the spectral function $\rho^{high}$ is
included. It is expected that as $\rho(\omega)$ decreases faster 
than $\omega^{-2}$ and due to the factor
, $cosh(\omega(t-\beta/2))/sinh(\omega \beta/2)$ in Eq.\ref{f-series},
the Matsubara Green's function
$G_{\beta}(t)$ will not be sensitive to the behavior of $\rho(\omega)$ 
in high $\omega$ region, except at $t=0$ and $t=\beta$.
We think that the $\eta$  and $\eta/s$ will not become 10 times of the
present value when more accurate determination of the transport
coefficients are carried out.

However it is important to make a more accurate
calculation of the transport coefficients, independent of the assumption
of the spectral function. For that goal, we are starting the calculation
of $G_{\beta}(t_n)$ 
on an anisotropic lattice, to apply the maximum entropy method.

\end{document}